\documentclass[a4paper,11pt]{article}
\pdfoutput=1 

\usepackage{jinstpub} 

\title{X-ray characterization of BUSARD chip: A HV-SOI monolithic particle detector with pixel sensors under the buried oxide}

\author[a,c,d,1]{Fabricio~Alcalde~Bessia,\note{Corresponding author.}}
\author[a,b,c,d]{Jos\'e~Lipovetzky,}
\author[e]{Ivan Peri\'c}

\affiliation[a]{Centro Atómico Bariloche and Instituto Balseiro,\\Av.E.Bustillo 9500, R8402AGP San Carlos de Bariloche, Argentina}
\affiliation[b]{Comisión Nacional de Energía Atómica (CNEA),\\Av.E.Bustillo 9500, R8402AGP San Carlos de Bariloche, Argentina}
\affiliation[c]{Universidad Nacional de Cuyo (UNCUYO),\\Av.E.Bustillo 9500, R8402AGP San Carlos de Bariloche, Argentina}
\affiliation[d]{Consejo Nacional de Investigaciones Científicas y Técnicas (CONICET),\\Av.E.Bustillo 9500, R8402AGP San Carlos de Bariloche, Argentina}
\affiliation[e]{Karlsruhe Institute of Technology (KIT),\\Hermann-von-Helmholtz-Platz 1, 76344 Eggenstein-Leopoldshafen, Germany}

\emailAdd{falcalde@ib.edu.ar}

\abstract{This work presents the design of BUSARD, an application 
specific integrated circuit (ASIC) for the detection of ionizing 
particles. The ASIC is a monolithic active pixel sensor which has been 
fabricated in a High-Voltage Silicon-On-Insulator (HV-SOI) process that 
allows the fabrication of a buried N+ diffusion below the Buried OXide 
(BOX) as a standard processing step. The first version of the chip, 
BUSARD-A, takes advantage of this buried diffusion as an ionizing 
particle sensor. It includes a small array of 13$\times$13 pixels, with 
a pitch of \SI{80}{\um}, and each pixel has one buried diffusion with a 
charge amplifier, discriminator with offset tuning and digital 
processing. The detector has several operation modes including particle 
counting and Time-over-Threshold (ToT). An initial X-ray 
characterization of the detector was carried out, obtaining several 
pulse height and ToT spectra, which then were used to perform the 
energy calibration of the device. The Molybdenum $\mathbf{K_{\alpha}}$ 
emission was measured with a standard deviation of \SI{127}{e^{-}} of 
ENC by using the analog pulse output, and with \SI{276}{e^{-}} of ENC 
by using the ToT digital output. The resolution in ToT mode is 
dominated by the pixel-to-pixel variation.
}


\graphicspath{{./img/}}
\DeclareGraphicsExtensions{.pdf,.jpeg,.png}

\newcommand{\code}[1]{\texttt{#1}}

\usepackage[detect-all]{siunitx}

\usepackage{tabu}
\usepackage{booktabs}

\usepackage[caption=false,font=footnotesize]{subfig}

\usepackage{url}

\begin{document}

\maketitle
\flushbottom

\section{Introduction}

Over the last decades the scientific community has been trying to 
develop more faster, accurate and low cost ionizing particle detectors, 
and for these purposes the CMOS technology has been the preferred 
choice because of its maturity and simple design flow. CMOS technology 
has been used for the development of readout integrated circuits in the 
so called \emph{hybrid} pixel detectors, in which the sensor is 
produced on a special fabrication process different from the readout 
chip. This type of approach has been explored and developed 
successfully since more than three decades ago and a history 
review can be found in \cite{heijne20011semiconductor}. Since then, 
hybrid detectors have been used for several experiments, like 
the LHC's ATLAS and CMS pixel detectors 
\cite{grosseknetter2005atlas,hugging2005front}, and they were used also 
by the \emph{Medipix} collaborations \cite{medipixcollaboration}. 
This collaboration developed hybrid detectors for medical imaging and 
high-energy particle physics experiments \cite{ballabriga2018asic}, like 
\emph{Medipix} and \emph{Timepix} detectors. 

However, hybrid detectors suffer from serious drawbacks in terms of 
costs associated with the assembly of hybrid parts. With the 
advancement of the CMOS consumer electronics and CMOS imaging 
applications, the development of \emph{Monolithic} Active Pixel Sensors 
(MAPS) rapidly emerge \cite{wermes2019pixel}. In a monolithic detector, 
the sensor and electronics are produced in the same piece of silicon as 
part of one single fabrication process 
and there is no need of additional post-processing steps, which can 
reduce the cost and failure rate in comparison with their hybrid 
counterpart. Since then, several MAPS have been designed that take 
advantage of different properties of CMOS technology: diffusion in the 
epitaxial layer \cite{dierickxB1997near}, high-voltage CMOS 
\cite{peric2007novel,peric2020high}, Partially-Depleted SOI (PD-SOI) 
\cite{bulgheroni2004monolithic}, Fully-Depleted SOI 
\cite{arai2010developments}, just to name a few. The main idea behind 
SOI MAPS is to make the particle-sensitive device in the handle 
wafer---possibly taking advantage of a high-resistivity substrate and 
thus a larger depletion region with a stronger electric field---and the 
active electronics in the top silicon film. The success of these 
devices was leveraged by the progress of the SOI technology, which made 
possible the fabrication of wafers with less defects both in the thin 
silicon film and in the handle wafer \cite{bulgheroni2004monolithic}. 
More recently, several works 
\cite{hemperek2015monolithic,fernandezperez2015radiation,fernandezperez2016charge,benka2018characterization,havranek2018maps,havranek2018x,marcisovska2020tid} 
presented the design and characterization of a MAPS on a \SI{180}{\nm} 
PD-SOI fabrication process which exploits the buried oxide to isolate 
the sensor device from the active electronics, connecting both through 
a hole in the BOX. These works include the description of 
the design and several tests like radiation hardness, charge collection 
in the sensitive junction, Total Ionizing Dose (TID) and even Single 
Event Effect (SEU) tolerance. 

Following the same principle of using a SOI fabrication process to make 
a monolithic sensor, in this work the BUSARD-A is presented. This chip 
is a prototype of a MAPS with the ability of counting particle hits and 
also measuring particle energy. It is intended to perform a basic 
energy discrimination of ionizing particles and energy resolved imaging 
on a future bigger array. Section \ref{sec:design} presents the design 
of the detector. A pixel cross-section is shown with the 
radiation-sensitive device and its connection to the active 
electronics. Then, the analog signal processing, digital blocks and 
general chip architecture are described. Next, section 
\ref{sec:xraycharacterization} presents the characterization of the 
device using the emission of fluorescence X-rays of different 
materials. BUSARD's analog output and and digital ToT data are used to 
obtain the energy spectra of the emitted photons. Then these spectra 
are used to trace energy calibration curves and to calculate the 
sensor's readout noise. The article finishes with a short conclusion 
highlighting the most relevant properties of the ASIC.

\section{Design of the sensor and builtin electronics} 
\label{sec:design}

The integrated circuit was designed and fabricated using the partially 
depleted Silicon-On-Insulator (SOI) \SI{180}{\nm} fabrication process from 
XFAB XT018. This process allows the creation of an 
n-type diffusion in the silicon substrate under the buried oxide and 
allows the connection of this buried diffusion to the active 
electronics located on the top silicon film. 

\begin{figure}
    \centering{}
    \includegraphics[width=0.8\linewidth]{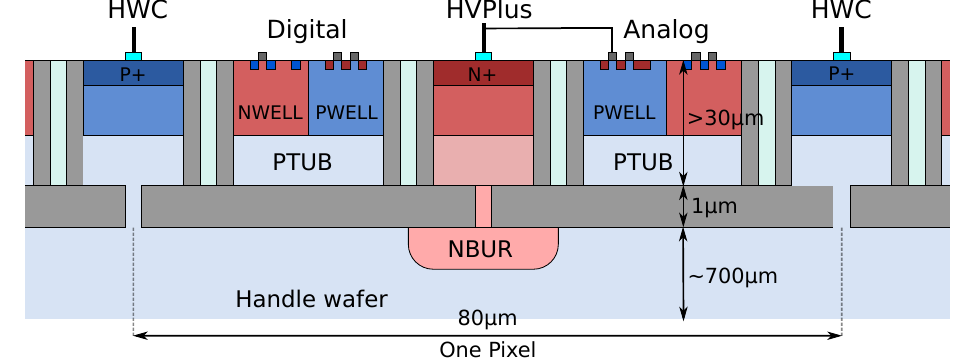}
    \caption{Pixel cross section. The fabrication process allows to 
    make a diode in the handle wafer and to contact it from the top 
    thin-silicon film. The buried diffusion is surrounded by the analog 
    and digital parts in the silicon film, which are isolated by deep 
    trench isolation.}
    \label{fig:PixelCrossection}
\end{figure}

Figure \ref{fig:PixelCrossection} shows a pixel cross section. The 
NBUR/p-subs device is the radiation sensitive element. The charge 
generated by impinging ionizing radiation is collected in its depletion 
region, which extends mostly on the p side of the junction, it flows to 
the device terminals and it is amplified by a charge amplifier located 
in the top silicon film. The NBUR/p-subs diode can withstand up to 
\SI{110}{\volt} applied in reverse without entering avalanche breakdown 
mode. This allows a large depletion region and thus a larger charge 
collection volume than standard low-voltage CMOS technologies. With the 
standard handle wafer resistance, the extension of depletion region 
into the p substrate gives approximately \SI{33}{\um} at \SI{110}{\V}.

The pixel also comprises one analog and one digital part, which are 
isolated from each other and from the NBUR connection using Deep Trench 
Isolation (DTI) isolation. 

\subsection{Pixel analog subsystem}

The analog subsystem is comprised by three cascaded stages: a first 
stage which is a charge amplifier, a second summing stage, and a 
comparator/discriminator stage. The buried N diffusion is biased through 
a diode-connected standard PMOS transistor Tbias built in its own pocket, so 
applying a high voltage to this device does not affect the rest of the 
electronics. 

The first stage is responsible for the amplification and translation 
from charge to voltage pulse. The conversion factor is \SI{6.3}{\uV} 
per e$^{-}$ of charge collected in the buried N diffusion, and, as will 
be shown later, it was measured by fitting the spectrum of several know 
radiation sources. The feedback capacitor, Cfb$_1$, has a MOS transistor in 
parallel, Tfb$_1$, biased with a voltage in such a way that most of the time is 
off, and during a pulse it will discharge the capacitor at a 
programmable rate allowing the adjustment of the tail length. Table 
\ref{tab:busardkeyparam} show the values of the capacitors and some 
key parameters of this stage. 

\begin{figure}
    \centering{}
    \includegraphics[width=0.8\linewidth]{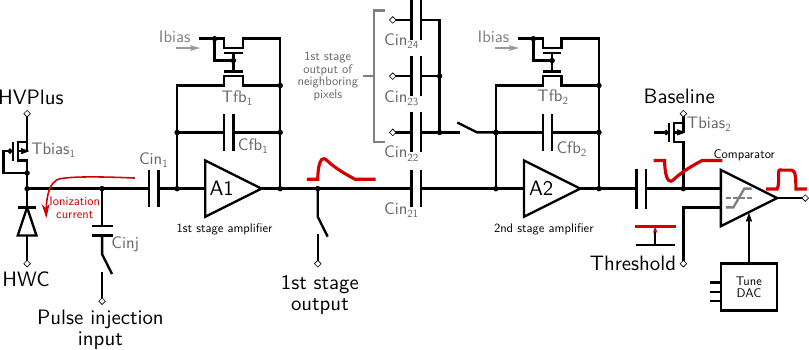}
    \caption{Schematic of the analog part of the pixel.}
    \label{fig:pixelAnalogAll}
\end{figure}

The purpose of the second stage is to sum the pulses produced by the 
first stage of neighboring pixels and, in this way, perform an analog 
binning, avoiding the loss of charge due to incomplete charge collection. 
This function can be enabled or disabled at any time with a configuration 
bit. When disabled, the pulse of the first stage passes unattenuated to 
the comparator stage. 

The output of the second stage is AC coupled to the comparator and 
this allows tuning the \emph{baseline} DC level, which is set by a voltage 
generated outside the chip. Then the comparator/discriminator compares 
this level with the \emph{threshold} voltage, also generated outside the 
chip. A digital pulse is generated and maintained each time the analog 
signal drops below the threshold voltage. Both the baseline and the 
threshold voltages are common and distributed to the whole pixel matrix.

The comparator offset can be adjusted by means of a Tune DAC circuit 
located in every pixel. This circuit is a 3-bit current steering DAC 
that introduces an unbalance into the differential amplifier of the 
comparator proportional to the digital word set.

In order to test the device there is a \emph{test pulse input} that is 
very useful for testing, but also for calibration of the whole 
matrix. A memory bit on each pixel allows opening or closing the switch 
that enables/disables this function. A voltage step 
injected to this input produces certain amount of charge to be 
transferred to the input of the first stage, producing a signal similar 
to the one produced by a particle interaction. 

It is also possible to route off chip the output of the 1st stage 
amplifier of one pixel in order to measure its response. This was made 
possible only on the first column of the matrix in order to reduce the 
layout complexity.

\subsection{Pixel digital subsystem}

\begin{figure}
    \centering{}
    \includegraphics[width=0.7\linewidth]{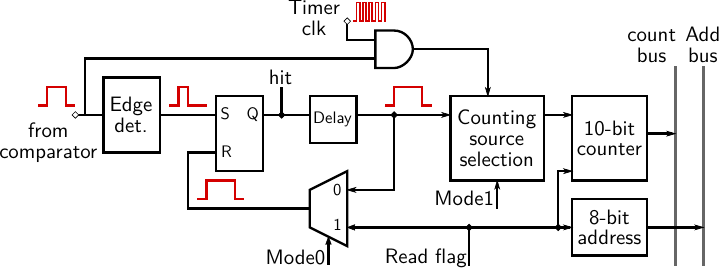}
    \caption{Block diagram of the digital subsystem inside each pixel.}
    \label{fig:pixelDigitalsubsystem}
\end{figure}

The digital subsystem inside each pixel is depicted in figure 
\ref{fig:pixelDigitalsubsystem}. The output of the analog comparator 
passes through an edge detector generating a fast pulse that sets an SR 
flip-flop, rising the \code{hit} signal. Here there are four modes of 
operation: (1) standard particle counting, (2) hit mode, (3) 
time-over-threshold (ToT), and (4) hit mode plus TOT. In the particle 
counting mode, the same \code{hit} signal is used to reset the 
flip-flop. A short configurable delay---implemented as a MOS current 
source and a capacitor--- was added to assure the proper operation and 
avoid racing conditions or missing counts. In this mode the same 
\code{hit} pulse is used to increment a 10-bit counter, so each time a 
particle hits the sensor the count gets increased by one. When a read 
operation begins, the counter is blocked and its tristate output is 
connected to the count bus, which is shared across all pixels. There is 
also a hardcoded address block in each pixel and a shared address bus, 
and each time a read operation is performed the pixel that is being 
read takes both buses.

In the second mode (hit mode) the difference is that the flip-flop is 
not reset by its own generated hit signal, but it is reset by a read 
operation instead. So, when particle interaction is detected the 
counter increases by one but then the pixel gets locked waiting until a 
read operation is performed.

Modes (3) and (4) are similar to (1) and (2), respectively, but in this 
case the counter increment is done with a fast clock signal generated 
of chip and distributed to the whole matrix. The count keeps increasing 
for as long as the comparator is in a high state, which is the same 
time as the pulse is below the threshold voltage. The result is a count 
that is proportional to the time the signal has been below the 
threshold and this, in turn, is proportional to the charge produced by 
the particle interaction.

In order to select a pixel for reading out its counter and address 
there are two possibilities: direct addressing and scan chain 
selection. In direct addressing the pixel is selected by setting a $1$ 
in the desired position of the row and column selection registers. 
These registers are 13 bits long, so each bit corresponds to a row or 
column and the pixel that has both set is selected. 

\begin{figure}
    \centering
    \includegraphics[width=0.7\linewidth]{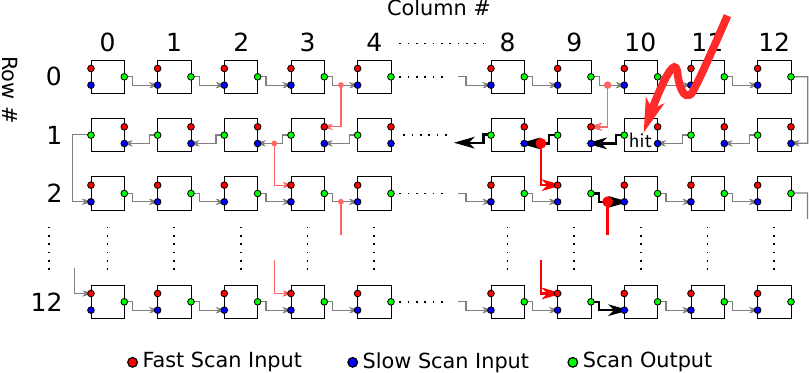}
    \caption{Scan chain logic connection. All pixels are connected in 
    a zig-zag slow chain with two fast propagation lines in columns 3 
    and 9. The thicker lines illustrate how the signal is propagated 
    through the fast chain. All floating inputs are connected to ground.}
    \label{fig:scanChain}
\end{figure}

The scan chain logic allows the automatic selection of the 
pixel to be read \cite{peric2020high,zhang2001development}. Pixels are connected together in a 
daisy chain array by using a special logic and the pixel that has been 
hit by a particle takes control of the data buses, locking access to 
every other pixel down the chain. A diagram of this chain is shown in 
figure \ref{fig:scanChain}.

There are two inputs \code{slowScanI} and \code{fastScanI}, and one 
output \code{outScan}. When a pixel is marked as hit (\code{hit=1}), 
the \code{outScan} output will be set. The next pixel in the chain will 
receive a `1' in its \code{slowScanI} input and this prevents this 
pixel from being read. This signal will propagate to the \code{outScan} 
of the second pixel and then to the third pixel down the chain, and so 
on. This is called the \emph{slow} scan chain because it passes through 
every pixel, therefore the signal delay from the hit position to the 
last pixel in the chain can be significant.

In order to reduce this delay time, two \emph{fast} propagation lines 
were made in columns 3 and 9. The \code{outScan} of pixels in those 
columns is connected to pixels of the next row. In this way the 
information arrives faster to the next rows and there is a tree-like 
propagation. The worst case scenario is when a particle hits the pixel 
located in first row and first column. In this case, in order for the 
information to get to the last pixel in the chain the signal must pass 
through 26 pixels.

Only when a pixel is selected for reading, by either using direct 
addressing or the scan chain, its counter can be cleared with a reset 
signal.

\subsection{General chip design and layout}

\begin{figure}
    \centering
    \includegraphics[width=0.5\linewidth]{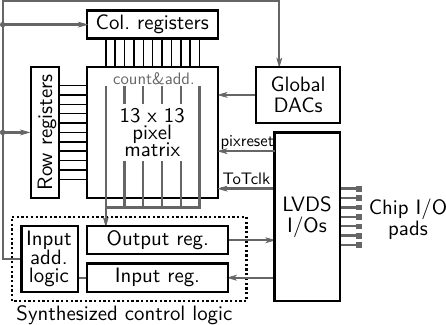}
    \caption{Diagram showing the chip architecture. There is 
    a digital synthesized logic that controls the configuration input 
    and readout of the matrix.}
    \label{fig:detectorDiagram}
\end{figure}

It is a small test chip comprising a matrix of 13 by 13 square 
pixels, each with a lateral size of \SI{80}{\um}. Figure 
\ref{fig:detectorDiagram} shows a diagram of the chip architecture. The 
pixel array has several row and column configuration registers whose 
functions are pixel addressing, loading tune-DACs, enabling the test 
pulse input for one or several pixels and enabling the 1st stage output. 
There are also 14 global 6-bit DACs whose purpose is to 
give some flexibility to the analog biasing of the pixel matrix. With 
the global DACs it is possible to fine tune the biasing of the first 
and second stage amplifiers and also to adjust the biasing of the MOS 
resistive feedback. In addition, the comparator DC current and Tune-DAC 
current reference can be tuned separately, allowing the calibration of 
the tune-DAC LSB step, and thus the threshold offset, with more 
precision. The digital 6-bit word of each DAC is stored in its own 
memory register. 
    
In order to load a register, the desired value and the register address 
are loaded in series into the input register by using a data and clock 
input lines. After receiving 18 bits, the control logic decodes the 
input and stores the value in the addressed register (row, column, 
configuration or DAC register).

The matrix is readout one pixel at a time by pulsing a clock line an 
reading the data coming out in series of the output register. This 
register contains the 10 bit count and 8 bit address of the pixel that 
is being read out. Its content must be loaded from one of the pixels in 
the matrix before reading it. In order to select a pixel---or several 
pixels---for reading, there is a special line called \code{pixHitLoad}. 
A pulse applied to this signal loads into the scan chain the pixels 
flagged as \emph{hit} (in hit mode), or pixels selected by direct 
addressing using the row and column selection registers. Then, a pulse 
applied to a second line called \code{pixRead} transfers the count and 
address from the pixel already set in the scan chain to the output 
register, also clearing its hit flag and allowing the read of the next 
in the chain. After reading out the output register the sequence 
repeats until the desired number of pixels is accessed. The advantage 
of this system is that it is not necessary to read the whole matrix 
every time, but instead just the pixels that have been hit of flagged 
for reading.

The time needed for reading one pixel is the time required to send the 
signals \code{pixHitLoad} and \code{pixRead} plus the time it takes 
reading out in series 18 bits from the output register. According to 
simulation results, the output clock can run as fast as \SI{200}{\MHz} 
and the total time for reading a pixel at maximum speed is 
\SI{100}{\ns}. However, while the output register is being read, the 
whole matrix---even the pixel whose count and address were stored in 
the output register---is active and prepared to detect ionization 
events. The only dead detection time introduced by this readout system 
is when the \code{pixRead} signal is high, which halts the pixel's 
counter to avoid data loss.

\begin{figure}
    \centering
    \includegraphics[width=0.9\linewidth]{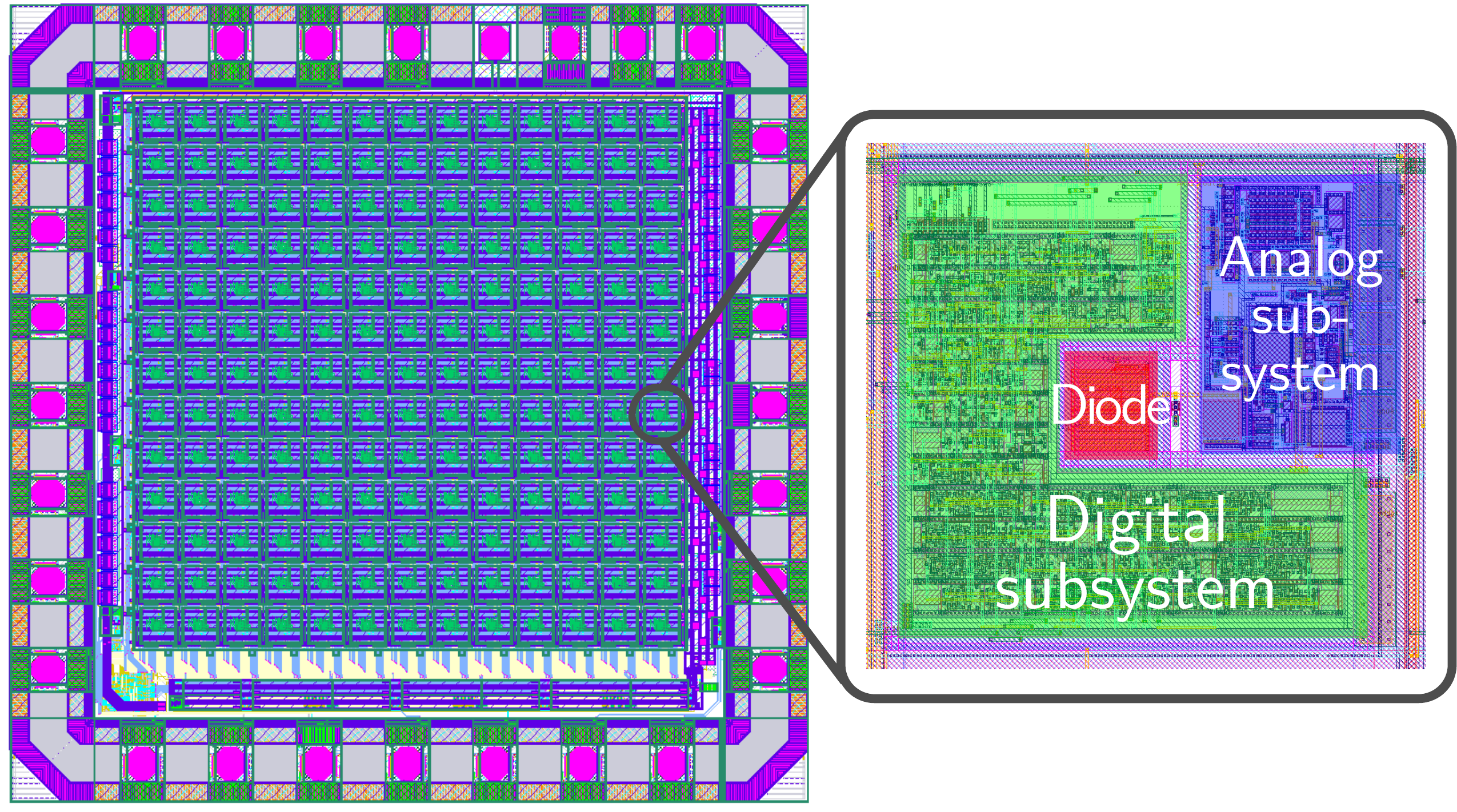}    
    \caption{Layout of the integrated circuit with a lateral size of 
    \SI{1.5}{\mm}. Most of the area is used by the $13\times13$ 
    pixel matrix, with a pixel pitch of \SI{80}{\um}. The global DACs are 
    located below the matrix and the synthesized digital controller is 
    fabricated in a \SI{180}{\nm} PD-SOI fabrication process. A detailed 
    view of one pixel is also shown.}
    \label{fig:layout}
\end{figure}

Figure \ref{fig:layout} shows the layout of the fabricated device. The 
13 by 13 pixel matrix takes most of the area of the \SI{1.5}{\mm} chip. 
The digital controller and global DACs are located below the matrix and 
the LVDS drivers near their output pads. All the digital input and 
output lines are Low-Voltage Differential Signals (LVDS) in order to 
reduce the digital noise coupled to the I/O ring.

Table \ref{tab:busardkeyparam} show some key parameters of the BUSARD-A
chip. 

\begin{table}
    \centering{}
    \begin{tabu} to 1\linewidth {l c l l}
            \toprule
            Parameter & Value & Unit & Comments\\
            \midrule
            $\mathrm{C_{det}}$ & $\approx1$ & \si{\fF} & \\
            $\mathrm{C_{in1}, C_{in2x}, C_{fb2}}$ & 50 & \si{\fF} & ~\\
            $\mathrm{C_{fb1}}$ & 5 & \si{\fF} & ~\\
            1st stage amp. peaking time & 1 to 5 & \si{\us} & Var. with 6-bit DAC for Ibias Tfb$_{1}$\\
            1st stage amp. discharge rate & 50 to 4.25 & \si{\mV\per\us} & Var. with 6-bit DAC for Ibias Tfb$_{1}$\\
            1st stage amp. power consumption & 10 & \si{\uW} & ~\\
            2st stage amp. power consumption & 18 & \si{\uW} & ~\\
            Comparator power consumption & $1.8$ & \si{\uW} & ~\\
            Baseline voltage & 900 & \si{\mV} & Generated outside the chip\\
            Threshold voltage & 885 & \si{\mV} & Generated outside the chip\\
            Pixel size & $80\times80$ & \si{\um} & ~\\
            Pixel power consumption & 30 & \si{\uW} & Only accounting for static power\\
            Matrix size & $13\times13$ & pixels & ~\\
            Chip power consumption & 20 & \si{\mW} & Mainly consumed by I/O drivers\\
            \bottomrule
    \end{tabu}
    \caption{BUSARD-A key parameters.}
    \label{tab:busardkeyparam}
\end{table}

\subsection{Pixel response calibration procedure}
\label{sec:pixelresponsecalibration}

Due to process variations and gradients it is expected to have a 
different response to the same input on each pixel. An equal pixel 
response is specially relevant when using the ASIC in ToT mode, in which 
a count proportional to the particle energy is expected. Differences 
come from the amplifiers gain, the feedback current sources and the 
comparators offset. In order to simplify the analysis, all these 
deviation sources can be modeled as a variation in gain and offset in 
the response of each pixel, according to the following equation

\begin{equation}
    y_{i} = m_{i} x + b_{i},
\end{equation}

where $x$ is the stimulus, $m_{i}$ and $b_{i}$ are the gain and offset 
respectively, and $y_{i}$ is the pixel's output. For the purpose of 
generating a stimulus in an easy way is that the test pulse input was 
added to the chip.

The calibration is based in the calculation of two arrays of constants 
$K_{1}$ and $K_{2}$ that can be used to equalize the pixel's output to 
the same mean value 

\begin{equation}
    \label{eq:calibrationformula}
    \bar{y} = K_{1,i} y_{i} + K_{2,i},\quad
    \left\{
    \begin{array}{lr}
        K_{1,i} = \frac{\bar{m}}{m_{i}}&~\\
        K_{2,i} = \bar{b} - K_{1} b_{i}&~
	\end{array}
    \right.
\end{equation}

The first step in the calibration procedure is to measure the response 
of every pixel to the same variable stimulus, this means applying test 
pulses of different amplitudes through the test-pulse-input, and 
extracting the average output $\bar{y}$, the average gain $\bar{m}$, 
and the average offset $\bar{b}$. Then, the calibration constant arrays 
are calculated as shown in equation \ref{eq:calibrationformula}. The 
length of these arrays is equal to the amount of pixels in the matrix. 
Finally, the calibrated mean output is obtained by multiplying the 
output of each pixel by its corresponding constant in the $K_{1}$ array 
and then adding the corresponding constant of the $K_{2}$ array.

\section{Low energy X-rays characterization}
\label{sec:xraycharacterization}

\begin{figure}
    \centering
    \includegraphics[width=0.7\linewidth]{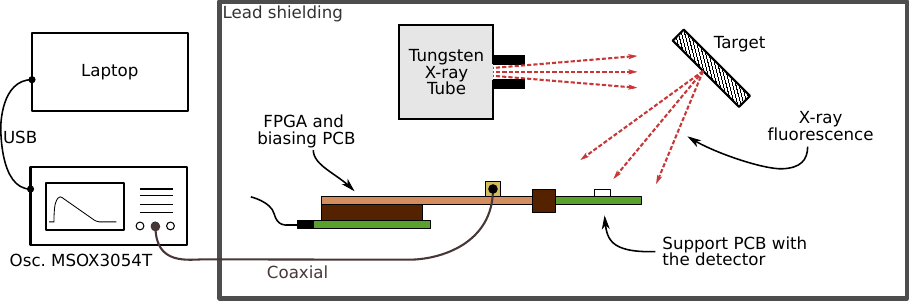}
    \caption{X-rays fluorescence experimental setup.}
    \label{fig:fluorescencesetup}
\end{figure}

The detector response was characterized by using low energy 
fluorescence X-rays. The setup shown schematically in figure 
\ref{fig:fluorescencesetup} was used to produce the fluorescence of 
different materials. These elements along with their most significant 
x-ray emissions are listed in table 
\ref{tab:fluorescenceXrayEmissions}. The sensor junction was biased at 
\SI{60}{\V}.

\begin{table}
    \centering{}
    \begin{tabu} to 1\linewidth {l S[table-format=2.2]}
            \toprule{}
            Z-Mat shell & [\si{\kilo\electronvolt}]\\
            \midrule{}%
            29-Cu \(K_{\alpha{}1}\) & 8.04\\
            30-Zn \(K_{\alpha{}1}\) & 8.63\\
            29-Cu \(K_{\beta{}1}\)  & 8.90\\
            30-Zn \(K_{\beta{}1}\)  & 9.57\\
            82-Pb \(L_{\alpha{}1}\) & 10.55\\
            82-Pb \(L_{\beta{}1}\)  & 12.61\\
            40-Zr \(K_{\alpha{}1}\) & 15.78\\
            42-Mo \(K_{\alpha{}1}\) & 17.47\\
            40-Zr \(K_{\beta{}1}\)  & 17.67\\
            \bottomrule
    \end{tabu}
    \hspace{2em}
    \begin{tabu} to 1\linewidth {l S[table-format=2.2]}
            \toprule{}
            Z-Mat shell & [\si{\kilo\electronvolt}]\\
            \midrule{}%
            42-Mo \(K_{\beta{}1}\)  & 19.60\\
            47-Ag \(K_{\alpha{}1}\) & 22.16\\
            48-Cd \(K_{\alpha{}1}\) & 23.07\\
            49-In \(K_{\alpha{}1}\) & 24.20\\
            47-Ag \(K_{\beta{}1}\)  & 24.94\\
            50-Sn \(K_{\alpha{}1}\) & 25.27\\
            48-Cd \(K_{\beta{}1}\)  & 26.09\\
            49-In \(K_{\beta{}1}\)  & 27.27\\
            50-Sn \(K_{\beta{}1}\)  & 28.48\\
            \bottomrule
    \end{tabu}
    \caption{Most intense emission lines of the elements used in the 
    detector characterization \cite{thompson2009x}.}
    \label{tab:fluorescenceXrayEmissions}
\end{table}

First, the response of the first stage amplifier of one pixel located 
on the leftmost column of the matrix was measured (this column is the 
only one with analog outputs available). The analog output of this 
amplifier, shown in figure \ref{fig:pixelAnalogAll}, was connected to 
an oscilloscope with its threshold level set to capture the output 
pulses. In this way, each photon impinging the sensor produces an 
output pulse with a height proportional to the ionization charge 
generated and collected in the sensor junction. Then, the signal 
waveforms were transferred to a laptop and stored for post-processing. 
The result is shown in section \ref{sec:firststagepulseheight}.

In another experiment, maintaining the same setup, the digital readout 
was used in order to acquire data. Using an FPGA, an USB connection 
to a computer, and a custom software, the detector was configured in 
ToT mode and the 169 pixels were read while the sensor was exposed to 
the x-ray fluorescence photons. In this case, a counter value different 
than zero means that the pixel was hit, and that value multiplied by the 
ToT clock period is the time that the pulse generated by the particle 
was above the threshold. On the other hand, if the pixel count is zero, 
that means that the pixel was not hit, so it can be skipped. The whole 
matrix was read in successive operations, storing the digital TOT per 
pixel when it was different than zero, and as many times as needed to 
acquire enough data. Section \ref{sec:totxrays} shows the result of 
this second experiment.

\subsection{Pulse height spectra}
\label{sec:firststagepulseheight}

\begin{figure}
    \centering
    \includegraphics[width=0.6\linewidth]{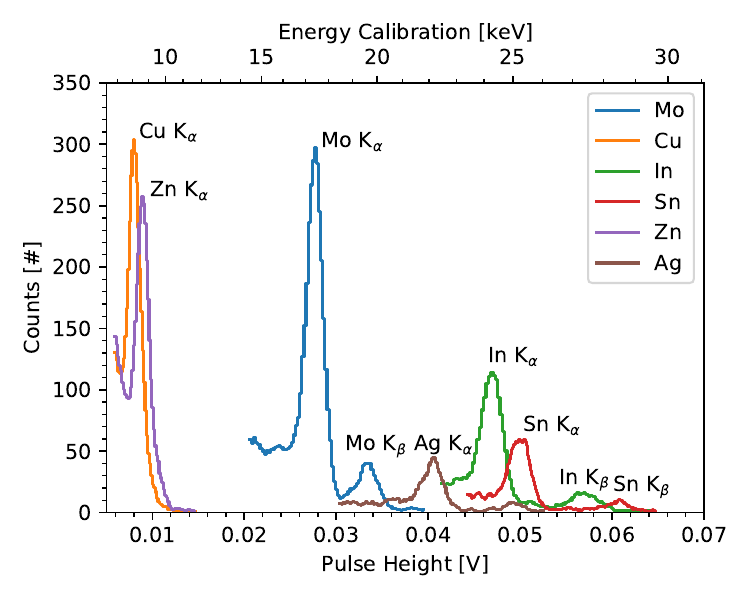}
    \caption[]{
    Measurement of the pulse height histogram for each of the elements 
    used in the experiment. They were obtained by monitoring the output 
    voltage of the first stage of a single pixel located at the edge of 
    the matrix. Only the peak part is shown for clarity. The top energy 
    axis was calculated according to the energy calibration of figure 
    \ref{fig:pulseheightenergycalibrationcurve}.}
    \label{fig:pulseheightspectra}
\end{figure}

As it was mentioned before, the output of the first stage amplifier was 
connected to an oscilloscope and each time a photon produced ionization 
the output pulse was digitalized and stored. Hundreds of waveforms were 
recorded and then analyzed with a Python script. The analysis consisted 
in looking for the baseline of the pulse by averaging the samples 
before the rising edge and then measuring the pulse height as the 
difference between the waveform maximum voltage and the baseline level. 
Figure \ref{fig:pulseheightspectra} shows the calculated pulse height 
histograms for each of the target elements. 

\begin{figure}
    \centering
    \includegraphics[width=0.6\linewidth]{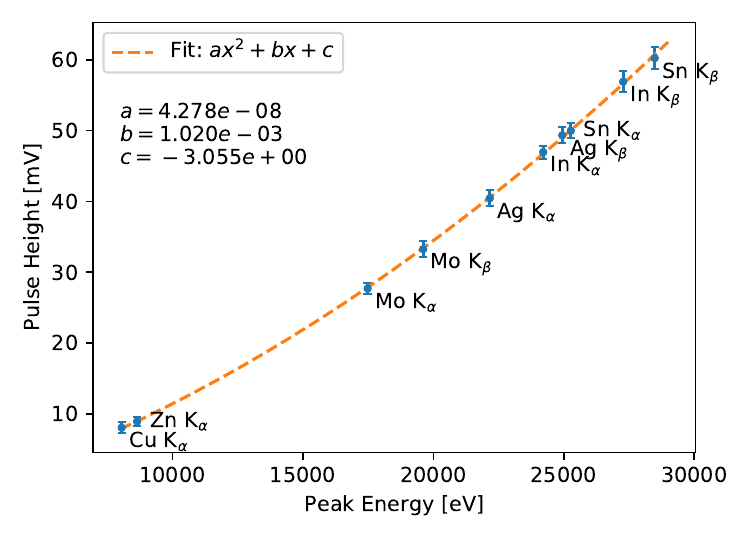}
    \caption[]{Pulse-height vs. energy curve. The energy of the 
    emission lines was taken from the table 
    \ref{tab:fluorescenceXrayEmissions} and its amplitude by fitting 
    the peaks of figure \ref{fig:pulseheightspectra} with a gaussian 
    curve and taking the mean value. The error bars represent the 
    standard deviation of the fits.}
    \label{fig:pulseheightenergycalibrationcurve}
\end{figure}

With this monolithic detector it was possible to distinguish the most 
intense emission lines of each of the elements used in the experiment, 
although the resolution does not allow to resolve between the copper 
$K_{\alpha1}$ and $K_{\beta1}$ emissions.

The emission peaks were fitted with a gaussian curve in order to 
extract their mean pulse height and standard deviation. These values 
were used in figure \ref{fig:pulseheightenergycalibrationcurve} to plot 
the pulse heights against the energy of the emission lines listed in 
table \ref{tab:fluorescenceXrayEmissions}, obtaining in this way the 
energy calibration curve. The response is slightly non-linear and so it 
was fitted with a quadratic function. The fitting function was then 
used to generate the top energy axis seen in figure 
\ref{fig:pulseheightspectra}.

The standard deviation of the Molybdenum $K_{\alpha}$ peak resulted in 
\SI{458}{\eV}, which can be translated to an Equivalent Noise Charge 
(ENC) in silicon of \SI{127}{e^{-}rms}. Due to the slightly non-linear 
response the gain is a bit higher at higher energies, so doing the same 
for the Indium $K_{\alpha}$ peak gives \SI{122}{e^{-}rms}. These 
numbers are similar to the standard deviation obtained in reference 
\cite{havranek2018x} for a $^{55}$Fe $\mathrm{K_{alpha}}$ peak 
(\SI{5.9}{\keV}), and slightly higher than the standard deviation 
obtained with a hybrid detector \cite{ballabriga2011characterization,
ballabriga2013medipix3}.

In order to measure the noise generated solely by the first stage 
amplifier, the X-ray source was shut off and the analog output was sampled 
with the oscilloscope at a rate of \SI{5}{Gs\per\s} during a time span 
of \SI{200}{\us}. The obtained signal has a gaussian noise distribution 
with a standard deviation of \SI{0.89}{\mV}, that is equivalent to 
\SI{100}{e^{-}rms} of noise charge, so most of the noise is introduced 
by the first stage amplifier.

It can be seen in figure \ref{fig:pulseheightspectra} that the peak 
have tails toward their left. These tails are produced by the 
incomplete charge collection (ICC) at the sensor diode. Photons 
ionizing the silicon below and outside the depletion zone produce 
charge that can diffuse and escape from the collection node, leading to 
smaller voltage pulses.

\subsection{Time-over-Threshold}
\label{sec:totxrays}

In this section the results of the measurements using the ToT mode are 
presented. The clock used for counting the ToT was set to 
\SI{200}{\MHz}, so a digital unit represents \SI{5}{\ns} in time. As 
before, the sensor junction was biased at \SI{60}{\volt}.

In this case, instead of reading just one pixel, ToT counts are 
acquired from the whole matrix, one value per pixel, and these values 
are used to create an X-ray spectrum by generating a histogram. In this 
case, the calibration of the sensor's response is needed due to the 
pixel to pixel variation and thus the procedure described in section 
\ref{sec:pixelresponsecalibration} was followed by using the test pulse 
input as stimulus. The baseline voltage was set to \SI{900}{\mV} and 
the threshold level to \SI{885}{\mV}, the second right below the voltage at which 
spurious counts are generated by noise. After calibration, triggering 
the sensor 4200 times with the same pulse gives a gaussian distribution 
of ToT with standard deviation of 4 units. Knowing that each unit 
represents \SI{5}{\ns}, this means that, for the same input, 68\% of 
the times the measured ToT will lay in a \SI{20}{\ns} interval around 
its mean value. The baseline and threshold voltages were maintained for 
all measurements.


\begin{figure}
    \centering{}
    \includegraphics[width=0.6\linewidth]{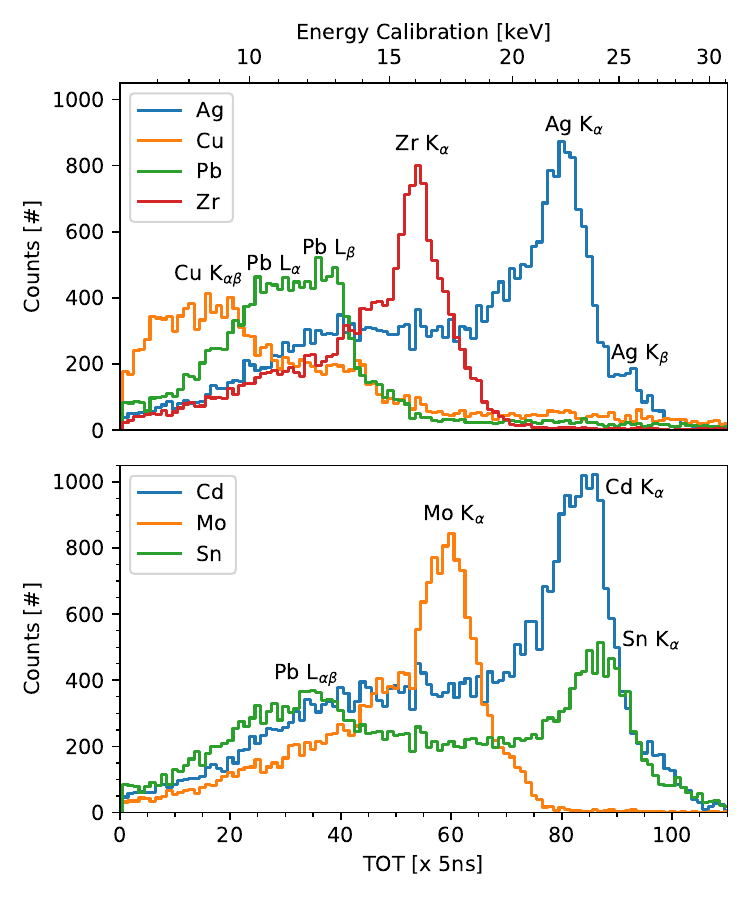}
    \caption[]{ToT histograms for each of the elements used in 
    the experiment. The top 
    energy axis was calculated according to the energy calibration of 
    figure \ref{fig:TOTcalibrationcurve}.}
    \label{fig:TOTspectra}
\end{figure}

Figure \ref{fig:TOTspectra} shows the obtained Time-over-Threshold 
histograms for the different target elements. The measurements were 
performed including ToT counts of the whole matrix after calibration. 
The principal emission lines are evident, although the energy 
resolution in this case is worse than using the pulse height of the 
first stage, as in figure \ref{fig:pulseheightspectra}. In this case 
the peaks have a long tail towards the left, effect that will be 
discussed in the next section. Also the spectrums of Cu and Sn have a 
notable contribution between 20 and 40 ToT counts that correspond to 
lead impurities. This is explained because the set of targets used in 
this case is different than that of section 
\ref{sec:firststagepulseheight}. 

\begin{figure}
    \centering
    \includegraphics[width=0.6\linewidth]{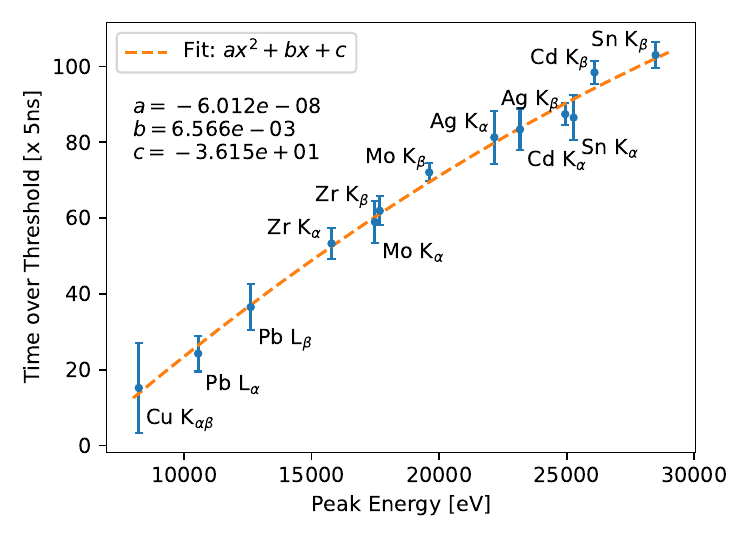}
    \caption[]{ToT vs. energy curve. The energy of the 
    emission lines was taken from the table 
    \ref{tab:fluorescenceXrayEmissions} and values in the y-axis where 
    obtained by fitting the peaks of figure \ref{fig:TOTspectra} with bimodal 
    curves. The error bars represent the standard deviation of the gaussian 
    model.}
    \label{fig:TOTcalibrationcurve}
\end{figure}

Again the pulses were fitted but this time using bimodal distributions 
in order to have a better fitting in the cases of mixed principal and 
secondary emissions. The mean and standard deviations along with the 
energy of the X-ray photons of table 
\ref{tab:fluorescenceXrayEmissions} were used to generate the 
calibration curve of figure \ref{fig:TOTcalibrationcurve}. The figure 
also shows a second order polynomial line that fits the data points. The 
coefficients obtained from this fit were used to generate the energy 
calibration top axis in figure \ref{fig:TOTspectra}. The ToT response 
is slightly non-linear and spans from \SI{8}{\keV} to \SI{30}{\keV} 
equivalent to 15 to 103 ToT counts, with a slightly better resolution 
for photons of higher energies. 

For the Mo peak, the standard deviation is $5.5$ ToT counts, equivalent 
to \SI{1}{\keV} or \SI{276}{e^{-}rms} of ENC. It was mentioned before 
that the response to the test pulse input has a standard deviation of 
4 ToT counts and this can be translated to \SI{200}{e^{-}rms} of ENC. So 
the pixel response calibration is the main contributor to the low 
energy resolution in the ToT mode.

The reason of the low number of ToT counts in figure 
\ref{fig:TOTspectra} is because the detector was designed for an energy 
range from \SI{8}{\keV} to \SI{100}{\keV}, so a big part of the 10-bit 
range was not used in these experiments. A faster clock would have 
extended the range, but it was not possible with the current setup and 
LVDS I/Os.

Another thing to notice in the ToT histograms of figure 
\ref{fig:TOTspectra} is that the peaks have a long tail towards the 
left. This is the same ICC effect that was explained in section 
\ref{sec:firststagepulseheight}.

\section{Conclusions}

With the monolithic and pixelated ionizing radiation detector presented 
in this work it was possible to count and measure the energy of X-ray 
photons. The fabrication process allows the fabrication of an n-type 
implant below the buried oxide that can be connected through the BOX 
with the active electronics. A reverse voltage of up to \SI{110}{\V} 
can be applied to the buried device, although for this work \SI{60}{\V} 
have been used. With the standard handle wafer resistance, the depletion 
region extends approximately \SI{25}{\um} into the p substrate at 
\SI{60}{\V} and this affects the charge collection efficiency of the 
device. The possibility of using a high-resistivity substrate will be 
studied in order to increase the depth of the depletion region.

The MAPS includes both analog and digital processing. It has been shown 
that the analog part is comprised by two cascaded amplifiers. The first 
stage has been characterized by using X-ray fluorescence photons and as 
part of the experiments the emission lines of several materials were 
obtained and distinguished. The emission peak of a Molybdenum target 
was obtained with a standard deviation of \SI{127}{e^{-}} of ENC and 
the noise introduced by the amplifier itself accounted for 
\SI{100}{e^{-}rms}. These numbers are slightly better than other works 
using the same technology \cite{havranek2018x}. 

Another characteristic of the analog electronics is the addition of a 
second stage whose purpose is to add the charge generated in neighboring 
pixels, although this operation mode has not been completely tested yet 
and will be reported in future works. 

The digital part has several operation modes that allow the counting of 
particles, the counting of the ToT and include the special \emph{hit} 
mode in which the pixel operation gets locked after detecting a 
particle hit until a read cycle has been performed. The matrix readout 
is governed by the scan chain and this reduces the amount of data 
transferred outside chip. This technique would significantly reduce the 
access time in large arrays. X-ray fluorescence spectra of various 
materials were obtained using the ToT mode with an ENC as high as 
\SI{276}{e^{-}}.

This ASIC has proven to be useful for the detection of X-ray photons 
and to perform a basic energy discrimination. Taking advantage of its 
energy and spatial resolution, future works will explore the 
possibility of using BUSARD covered with conversion layers for the 
detection of thermal and epithermal neutrons, a technique that was 
studied by using commercial CMOS image sensors \cite{perez2018thermal, 
perez2021high}.

\bibliographystyle{JHEP.bst}
\bibliography{Bibliografia}


\end{document}